\documentclass[aps,prb,twocolumn]{revtex4}
\usepackage{amsmath}
\usepackage{epsfig}
\usepackage{graphicx}
\usepackage{multirow}
\usepackage{array}
\usepackage{inputenc}
\newcommand{\rt}{R$_2$T$_2$O$_7$}
\newcommand{\tbti}{Tb$_2$Ti$_2$O$_7$}
\newcommand{\ybti}{Yb$_2$Ti$_2$O$_7$}
\newcommand{\ybsn}{Yb$_2$Sn$_2$O$_7$}
\newcommand{\gdsn}{Gd$_2$Sn$_2$O$_7$}
\newcommand{\erti}{Er$_2$Ti$_2$O$_7$}
\newcommand{\ersn}{Er$_2$Sn$_2$O$_7$}
\newcommand{\mub}{$\mu_{\rm B}$}

\newcommand{\er}{Er$^{3+}$}

\begin{document}
\author{Solene Guitteny$^{1}$, Sylvain Petit$^{1}$, Elsa Lhotel$^{2}$, 
Julien Robert$^{1}$, Pierre Bonville$^{3}$, Anne Forget$^{3}$, Isabelle Mirebeau$^{1}$}
\affiliation{$^1$ CEA, Centre de Saclay, DSM/IRAMIS/ Laboratoire L\'eon Brillouin, F-91191 Gif-sur-Yvette, France}
\affiliation{$^2$ Institut N\'eel, CNRS, 25 Av des martyrs, BP 25, 38042 Grenoble Cedex, France}
\affiliation{$^3$ CEA, Centre de Saclay, DSM/IRAMIS/ Service de Physique de l'Etat Condens\'e, F-91191 Gif-Sur-Yvette, France}

\title{Palmer-Chalker correlations in the XY pyrochlore antiferromagnet \ersn}
\date{\today}
\begin{abstract}
\ersn\, is considered, together with \erti, as a realization of the XY antiferromagnet on the pyrochlore lattice. We present magnetization measurements confirming that  \ersn\, does not order down to 100 mK but exhibits a freezing below 200 mK. Our neutron scattering experiments evidence the strong XY character of the \er moment and point out the existence of short range correlations in which the magnetic moments are in peculiar configurations, the Palmer-Chalker states, predicted theoretically for an XY pyrochlore antiferromagnet with dipolar interactions. Our estimation of the \ersn\, parameters confirm the role of the latter interactions on top of relatively weak and isotropic exchange couplings.

\end{abstract}

\maketitle

\section{Introduction}

Geometrical frustration has become a central challenge in contemporary condensed matter physics. It is 
the source of many exotic ground states whose description remains challenging for both theoreticians and 
experimentalists \cite{Lacroix}. These unconventional magnetic states often originate from the strong 
degeneracy of the ground state manifold, which prevents the stabilization of standard magnetic phases. 
Whatever their type, perturbations are often driving the low temperature behaviors by lifting partially or 
totally this extensive degeneracy. Quantum or thermal fluctuations may also enter into play to select and 
stabilize a particular configuration (or a subset of configurations), a phenomenon called "order 
by disorder" mechanism \cite{Villain80}. The family of pyrochlore compounds \rt\, (R is a rare earth and 
T=Ti, Sn, Zr, ...), with the rare-earth magnetic moments localized at the vertices of corner-sharing 
tetrahedra are model systems to study these subtle order by disorder effects \cite{gingrasrmp,bramgin00}. 

The case of R=Er compounds is of specific interest: they present a strong XY-like anisotropy, combined 
with antiferromagnetic interactions leading to a model with an extensive classical degeneracy \cite{champ,stasiak}. 
The easy magnetic planes are perpendicular to the local $<111>$ ternary axes (XY character), arising from 
the crystal field properties of the Kramers \er\, ion. While no signature of long-range order could be detected 
down to 100 mK in \ersn\, \cite{matsuhira, gardner, lago}, \erti\, undergoes a transition towards an antiferromagnetic 
N\'eel phase below $T_N=1.2$~K \cite{Blote69, Harris98, Siddharthan99}. This ordered phase has a non-collinear 
structure, in which the magnetic moments are perpendicular to the local $<111>$ axes in a peculiar configuration 
denoted $\psi_2$ \cite{champ2,poole}.
In \erti, this structure is surprising since dipolar interactions, which are an important perturbation to the isotropic 
exchange Hamiltonian, are expected to select other magnetic states, called Palmer-Chalker states \cite{palmer, champ2}. 
However, by considering general anisotropic exchange parameters, it has been recently argued that a quantum 
order by disorder mechanism \cite{clarty,zito,savary} explains this $\psi_2$ ordering and accounts for many 
experimental features. In this context, the reasons for the absence of ordering in \ersn\, remain an open question. 


In this paper, we address this issue by determining experimentally the key parameters of the Hamiltonian 
of \ersn: the crystal electric field (CEF) parameters obtained from inelastic neutron scattering experiments 
and the anisotropic exchange parameters deduced from the magnetization curves. The main difference 
between the titanate and stannate parameters is a weaker and less anisotropic exchange tensor. By 
analyzing neutron scattering data, we demonstrate 
the existence of short-range correlated domains frozen in the Palmer-Chalker configurations \cite{palmer}, 
hence quite different from the $\psi_2$ configuration selected in \erti\, \cite{champ2,poole}. We finally 
show that these configurations are indeed stabilized in a mean field calculation for this set of parameters. 

Magnetization and ac susceptibility measurements were performed on a powder sample down to 100 mK 
using a superconducting quantum interference device (SQUID) magnetometer equipped with a dilution 
refrigerator developed at the Institut N\'eel-CNRS Grenoble \cite{Paulsen01}. The neutron measurements 
were performed on the same sample at the cold triple-axis spectrometer 4F2 of LLB-Orph\' ee 
reactor \cite{neutron}. 


\begin{figure}
\centerline{
\includegraphics[width=7cm]{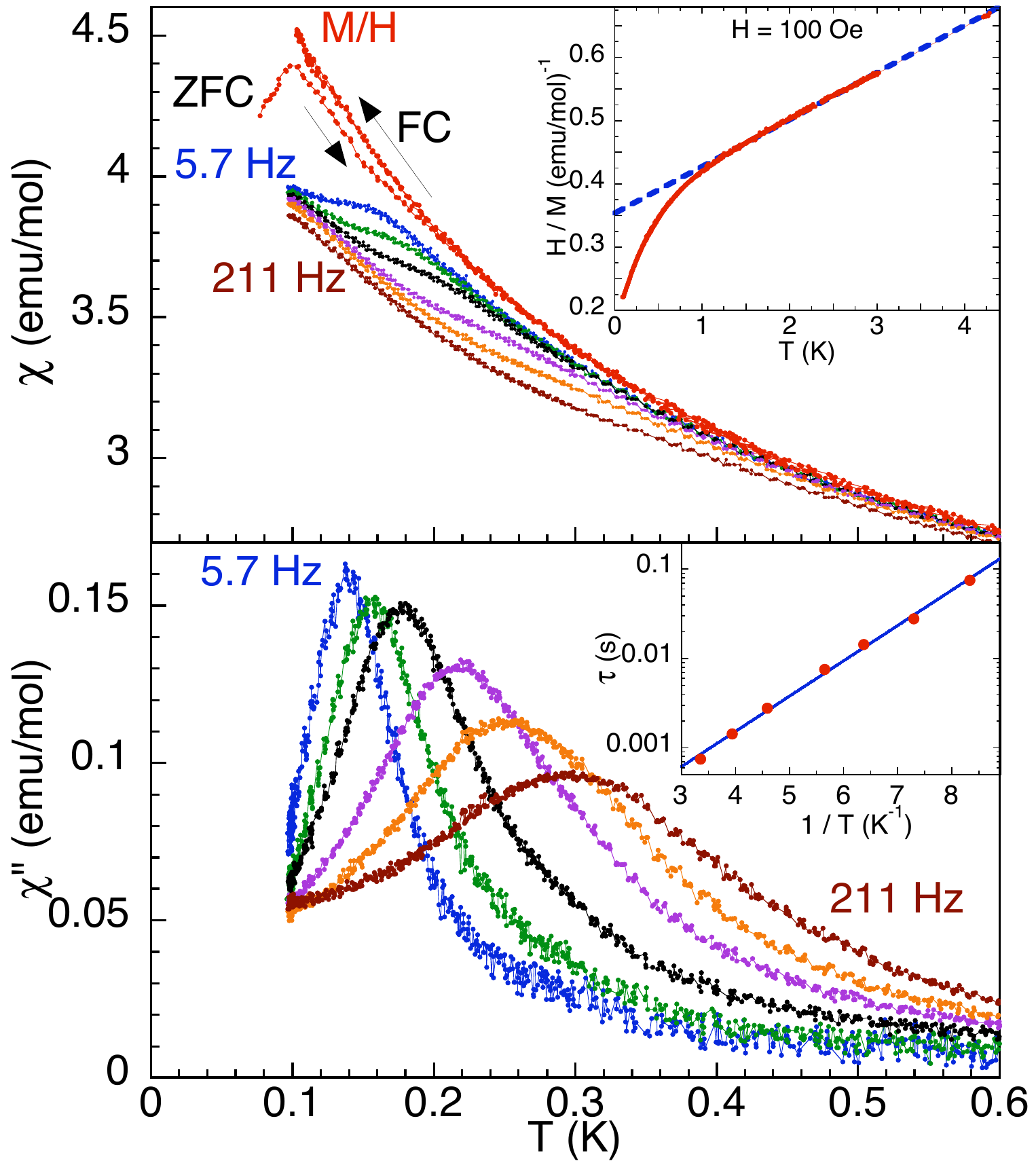}
}
\caption{(color online) ac and dc susceptibility vs temperature. Ac susceptibility is plotted for 
5.7 Hz$<f<211$ Hz with $H_{\rm ac}=1.4$ Oe. Top: $M/H$ and real part of the ac susceptibility 
$\chi'$ vs $T$. $M$ was measured using the ZFC-FC procedure with $H=50$ Oe. Inset: $H/M$ vs 
$T$ for $H=100$ Oe. The dotted line is a fit to the equation $H/M=0.35+0.07T$. Bottom: 
$\chi"$ vs $T$. Inset: $\tau$ vs $1/T_{\rm max}$ showing the Arrhenius behavior 
with $\tau_0=4 \times 10^{-5}$ s and $E/k_B=0.9$ K.} 
\label{figXT}
\end{figure}

\section{Magnetization and susceptibility}

Our magnetization measurements first confirm the absence of transition towards a long range ordered state
down to 100 mK. The dc susceptibility keeps increasing with decreasing temperature. It presents an upturn below 
about 2 K, hence deviating from a Curie-Weiss behavior (see the top inset of Figure \ref{figXT}), in agreement with 
Ref. \onlinecite{matsuhira}. Below 200 mK, a freezing is observed, as shown by an irreversibility in the zero field 
cooled - field cooled (ZFC-FC) magnetization and by a frequency dependence in the ac susceptibility (see top 
Figure \ref{figXT}). The imaginary part of the susceptibility $\chi"$ exhibits a peak 
whose frequency dependence can be accounted for by an Arrhenius law $\tau=\tau_0 \exp(E/k_BT)$ where 
$\tau=1/2\pi f$,  $\tau_0=4\times 10^{-5}$ s and $E/k_B=0.9$ K (see bottom Figure \ref{figXT}) in the 
measured frequency range (0.57 - 211 Hz). 
Qualitatively similar features have been observed in this temperature range in other spin-liquid compounds such 
as Gd$_3$Ga$_5$O$_{12}$ \cite{Schiffer95} or \tbti\, \cite{Gardner03, Hamaguchi04, Lhotel12}. At the moment, 
no clear picture emerges to explain this freezing, but it could be associated with slow dynamics of correlated spins. 

\begin{figure}
\centerline{
\includegraphics[width=7cm]{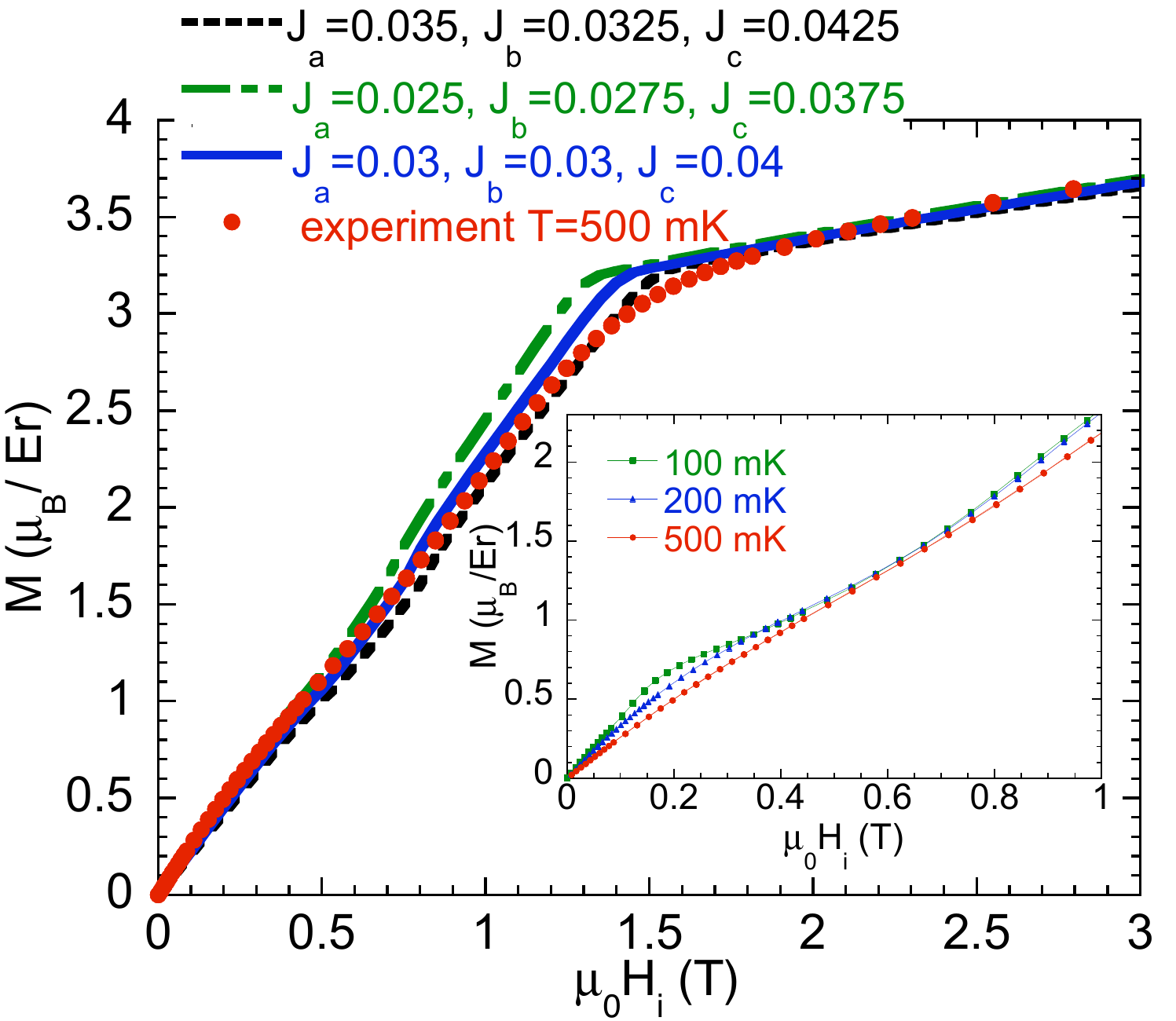}}
\caption{(color online) Magnetization $M$ vs internal field $H_i$ (points), along with simulation (lines) (see text).
Corrections for demagnetizing effects were made assuming a demagnetizing factor $N=4 \pi /3$ (cgs units) 
\cite{footnote1}. Results for different sets of parameters (assuming $J_4=0$) are presented to illustrate the 
sensitivity of the calculation. Inset: $M$ vs $H_i$ for $\mu_0H_i<1$ T at 100 (green squares), 200 (blue 
triangles) and 500 mK (red points). }
\label{fig2}
\end{figure}

The magnetization curves as a function of field present an inflection point around 1 T for temperatures below 
750 mK (see Figure \ref{fig2}). This behavior is reminiscent of the field induced transition observed in \erti\, 
\cite{ruff, cao, sosin, petrenko11}, thus suggesting that a field induced order might be stabilized above this 
field in \ersn. Unfortunately, the powder nature of the sample prevents from a detailed analysis of this 
metamagnetic like behavior. However, it is worth mentioning that preliminary calculations (using the 
mean-field model developed in Section \ref{discussion}) indicate that reorientations of the magnetic 
moments occur in the 1 - 1.5 T field range for the three main symmetry directions [110], [100] and [111]. 

Below 200 mK, an additional curvature develops in the magnetization curve around 0.2 T (see inset of Figure 
\ref{fig2}) which might be associated with the freezing observed in low field at these temperatures. 

\section{XY anisotropy and crystal field analysis}

\begin{figure}
\centerline{
\includegraphics[height=6cm]{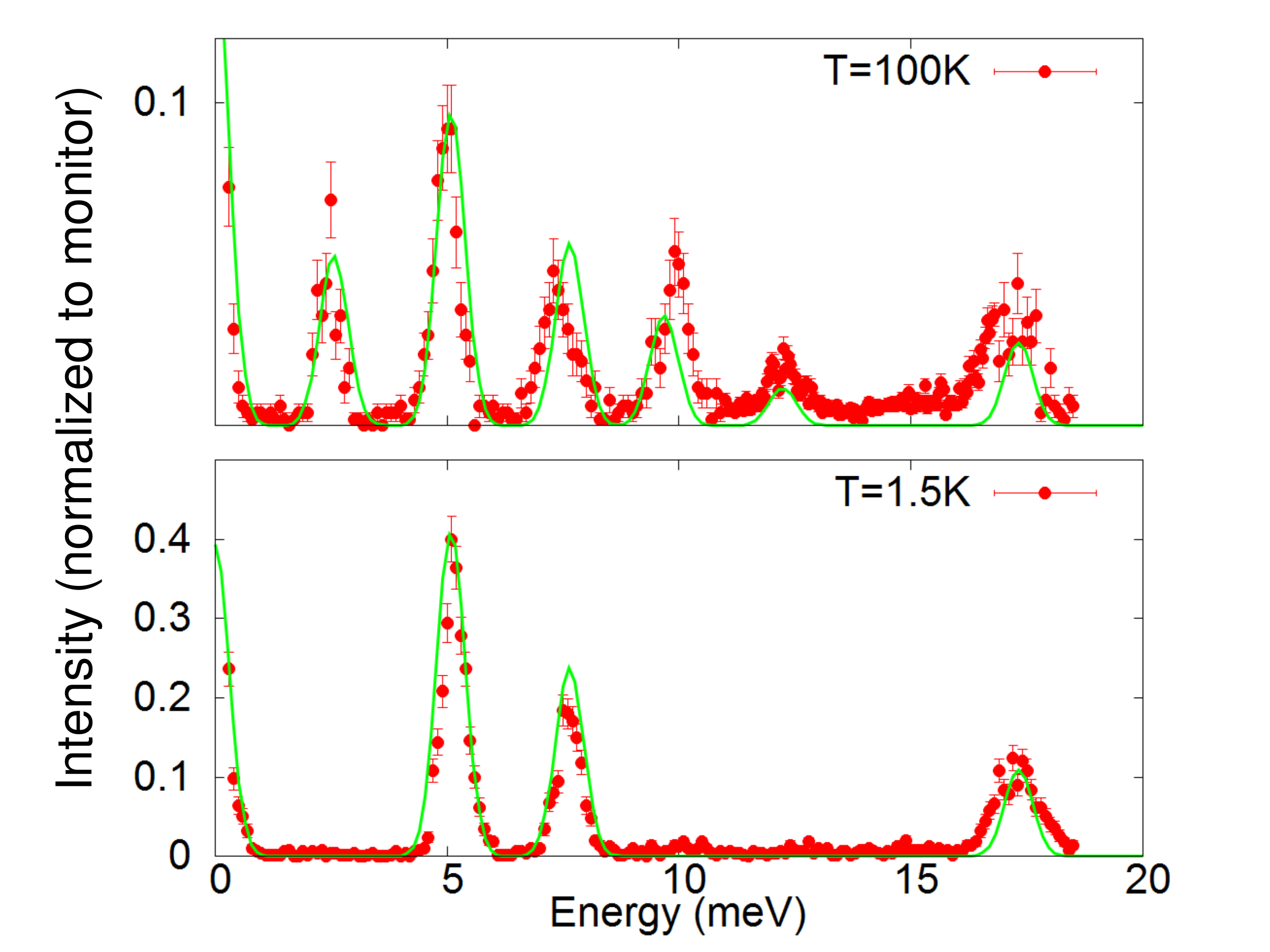}}
\caption{(color online)Inelastic neutron scattering spectra showing the CEF excitations. The lines 
correspond to the calculation (see text) with the parameters obtained from the fit at 1.5 K. At 100 K, 
the slight discrepancy is attributed to a small evolution of the parameters with temperature.
}
\label{fig3}
\end{figure}

Aiming at a precise determination of the \er\, anisotropy, the CEF excitations were measured by means of 
inelastic neutron scattering experiments (see Figure \ref{fig3}), carried out at temperatures of 1.5, 10, 50 
and 100 K. Between 0 and 20 meV, three CEF levels are observed at 
$E_1$ =5.1, $E_2$=7.6 and $E_3$=17.2 meV, in agreement with Ref \onlinecite{gardner}. With increasing 
temperature, excited states are populated to the detriment of the ground CEF state, giving rise to new 
modes at $\hbar \omega = E_i-E_j$. The analysis of these spectra is based on the simulation of the 
scattering function $S(Q,\omega)$: 
\begin{eqnarray*}
S(Q,\omega) = \sum_{m,n}\frac{e^{-E_m/k_BT}}{Z} 
\langle m | \vec{J} | n \rangle \langle n | \vec{J} | m \rangle \\
\times \delta(\omega+E_n-E_m)
\end{eqnarray*}
where the $|m\rangle$ and $E_m$ are respectively the eigen wavefunctions and eigenvalues of  
the CEF Hamiltonian ${\cal H}_{\mbox{CEF}}$ ($J=15/2$, $g_J=6/5$ for \er):
\begin{equation*}
{\cal H}_{\mbox{CEF}} = \sum_{m,n} B_{nm} O_{nm}
\end{equation*}
The $O_{nm}$ are the Stevens operators and the $B_{nm}$ the associated coefficients that remain 
to be determined (Ref. \onlinecite{cao2} and references therein). $Z$ is the partition function defined 
by $Z=\sum_m e^{-E_m/k_BT}$. Fitting the data  through this model yields the coefficients listed in 
table \ref{table-cef} (see also Appendix \ref{CEF}). The wavefunctions of the ground doublet lead to 
$g_{\perp}=7.52 \pm 0.1$ and $g_ {//}=0.054 \pm 0.02$. For comparison, the \erti\, values from 
Ref \onlinecite{cao2} are also given, showing that both compounds have rather similar CEF schemes, 
but that the \er\, magnetic moment has a stronger planar character in \ersn.

\begin{table}
\begin{tabular}{ccccccccccc}
\hline
   & B20 & B22 & B40 & B42 & B43 & B60 & B63 & B66 & $g_{\perp}$ & $g_{//}$ \\
\hline
\erti  & 616 & 0 & 2850 & 0 & 795 & 858 & -493 & 980 & 6.8 & 2.6\\
\ersn & 656 & 0 & 3010 & 0 & 755 & 738 & -653 & 990 & 7.52 & 0.054\\
\hline
\end{tabular}
\caption{Stevens coefficients (in K) for \ersn\,(present work) and \erti\, (from \cite{cao2}).}
\label{table-cef}
\end{table}


\begin{figure}
\centerline{
\includegraphics[height=6cm]{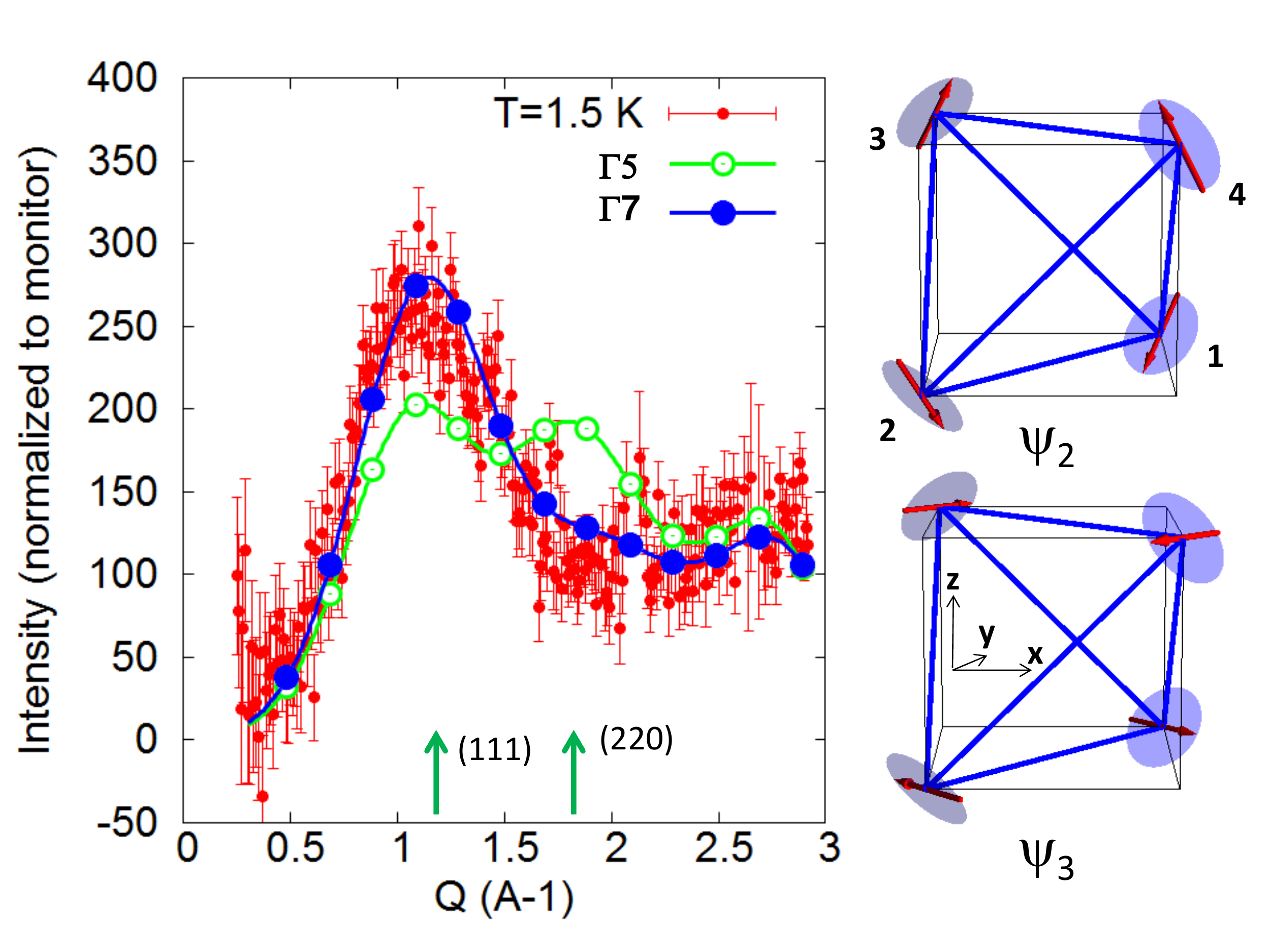}}
\caption{(color online)
(Left): Diffuse magnetic scattering measured by neutron scattering at 1.5 K. "High" temperature reference data 
($T$=50 K) have been subtracted. Exclusion zones have been considered around $Q$=2 $\AA^{-1}$ to eliminate an 
artifact due to a slight temperature shift of a nuclear peak. The lines are the result of a Rietveld fit assuming either 
the $\psi_{1,2} - \Gamma_5$ (green open circles) or the $\psi_{3,4,5} - \Gamma_7$ (blue solid circles) structure. 
(Right): Magnetic configurations $\psi_2$ and $\psi_4$ (see text and Table \ref{table-sym}) predicted by the 
symmetry analysis for the $\bf{k}$=0 propagation vector. 
}
\label{fig4}
\end{figure}


\section{Diffuse elastic scattering and Palmer-Chalker correlations}

To further describe the spin liquid state of \ersn, we have measured the spin-spin correlation function 
$S(Q,\omega)$ at 1.5 K. An elastic response is observed, forming a broad peak centered at $Q$=1.1 
$\AA^{-1}$, consistent with the results of Ref \onlinecite{gardner}. This response is typical of an 
elastic diffuse scattering where spin correlations extend over a few inter-atomic distances and are 
frozen at the time scale of the neutron probe. It is accompanied by a quasi-elastic contribution 
corresponding to fluctuations of this short range ordered pattern with typical rate $\gamma$ = 0.5 
meV, namely a typical time of $\tau \sim 10$ ps. The intensities of both contributions increase 
continuously with decreasing temperature. 

Such a diffuse peak does not preclude of any type of magnetic correlations in general. 
However, given the similarities between \ersn\, and \erti\,, both being antiferromagnets and sharing 
an XY anisotropy, we propose to model the magnetic ground state in \ersn\, by considering finite size 
magnetic domains (to account for the peak broadening), chosen among the symmetry allowed patterns 
for a $\bf{k}$=0 propagation vector. This modeling is based on a refinement which is constrained by 
symmetry and physical arguments, as explained below.

The symmetry analysis, performed in the space group Fd-3m using the BasIreps software 
\cite{Izyumov-Kovalev} shows that the basis states of the $\bf{k}$=0 manifold transform as linear 
combinations of the basis vectors of four Irreducible Representations (IR), labeled $\Gamma_{3,5,7,9}$ 
in group theory \cite{champ2,poole}. The XY anisotropy is minimized only for: i) linear combinations of the 
two basis vectors $\psi_1$ and $\psi_2$ which transform according to $\Gamma_5$; ii) a discrete set of basis 
vectors $\psi_{3,4,5}$ which transform according to $\Gamma_7$ \cite{notation}. 
The ground state of \erti\, and the Palmer-Chalker states (PC) \cite{palmer} correspond to $\psi_2$ and 
$\psi_{3,4,5}$ respectively, namely to different IR's. Table \ref{table-sym} and the right side of Figure 
\ref{fig4} provide the coordinates of these basis vectors and a sketch of the $\psi_2$ and $\psi_3$ 
magnetic structures (see also Appendix \ref{mag_struc}). 

We proceed by fitting the crystalline structure at 50K to determine 
the overall scaling factor and the lattice parameters. Using these values and assuming a 
given $\psi$ set, the two remaining parameters of the proposed model are the amplitude of the \er\,moment 
and the coherence length of the magnetic domains, which determines the width of the diffuse peaks. 
As shown in Figure \ref{fig4}, subtracting the high temperature data (50 K) to focus on the magnetic signal 
only, a very good refinement is obtained with the vectors $\psi_{3,4,5}$ of $\Gamma_7$, yielding an \er\, 
moment of 2.8 \mub\ at 1.5 K and a coherence length of about 10 \AA\, \cite{fullproff}. A much worse 
agreement is obtained with the vectors $\psi_1$ or $\psi_2$ of $\Gamma_5$. In the data ($Q<1.7 
\rm{~\AA^{-1}}$) of Ref. \onlinecite{gardner}, the diffuse peak, and so the \er\ moment, keep increasing 
down to 100 mK. By comparison with the present results, the \er moment likely reaches 3.8 \mub\ at 
100 mK. Note that powder measurements cannot distinguish between the basis vectors of either 
$\Gamma_5$ or $\Gamma_7$.
However,  for a given representation, the relative intensities of the (111) and (220) peaks are fixed. 
The choice between $\Gamma_5$ and $\Gamma_7$ IR is thus unambiguous.

\begin{table}
\begin{tabular}{cccccc}
\hline
&Site & 1 & 2 & 3 & 4 \\
&CEF axis & $(1,1,-1)$ & $(-1,-1,-1)$ & $(-1,1,1)$ & $(1,-1,1)$ \\
\hline
$\Gamma_5$ & $\psi_1$ & (-1,1,0) & (1,-1,0) & (1,1,0) & (-1,-1,0) \\ 
& & (0,1,1) & (0,-1,1), & (0,1,-1) & (0,-1,-1) \\
& & (1,0,1) & (-1,0,1) & (-1,0,-1) & (1,0,-1) \\
&$\psi_2$ & (1,1,2) & (-1,-1,2) & (-1,1,-2) & (1,-1,-2) \\
& & (-2,1,-1) & (2,-1,-1) & (2,1,1) & (-2,-1,1) \\
& & (-1,2,1) & (1,-2,1) & (1,2,-1) & (-1,-2,-1) \\
\hline
$\Gamma_7$ & $\psi_3$ & (1,-1,0) & (-1,1,0) & (1,1,0) & (-1,-1,0) \\
&$\psi_4$ & (0,1,1) & (0,1,-1) & (0,-1,1) & (0,-1,-1) \\
&$\psi_5$ & (-1,0,-1) & (-1,0,1) & (1,0,1) & (1,0,-1) \\
\hline
\end{tabular}
\caption{Coordinates of the moments at the 4 sites of a tetrahedron in the different $\psi$ sets (see text). 
Note that $\psi_{3,4,5}$ are obtained by reversing a pair of anti-parallel spins in the $\psi_1$ series.
}
\label{table-sym}
\end{table}

\section{Estimation of the exchange constants in \ersn\,, analysis and discussion}
\label{discussion}

We proceed with the estimation of the exchange constants in \ersn\, by combining neutron data 
and magnetization curve analyses. 
As emphasized above, applying a magnetic field drives the system towards an ordered state, 
hence making a mean field treatment an acceptable starting point. We thus follow the mean 
field approach proposed in Ref. \onlinecite{clarty}, and consider the Heisenberg Hamiltonian  
for R moments $\vec{J}_i$ at sites $i$ of the pyrochlore lattice: 
\begin{equation*}
{\cal H} = {\cal H}_{\mbox{CEF}} + 
\sum_{<i,j>} \vec{J}_i ({\cal \tilde J}+{\cal \tilde J}\mbox{dip}_{i,j})
\langle \vec{J}_j \rangle + g_J \mu_B \vec{H}. \vec{J}_i
\end{equation*}
In this expression, $\vec{H}$ is an applied magnetic field, ${\cal \tilde J}$ denotes an (anisotropic) 
exchange tensor and ${\cal \tilde J}\mbox{dip}_{i,j}$ the dipolar interaction limited to the contribution 
of the nearest neighbours. Various conventions have been used to define this anisotropic exchange
\cite{clarty,savary,zito,thompson,malkin}. Here, we assume an exchange tensor which is diagonal 
in the $(\vec{a},\vec{b},\vec{c})$ frame linked with a R-R bond \cite{malkin}:
\begin{eqnarray*}
\vec{J}_i {\cal \tilde J} \vec{J}_j &=& \sum_{\mu,\nu=x,y,z} J_i^{\mu} 
\left( 
{\cal J}_a a_{ij}^{\mu} a_{ij}^{\nu} +  
{\cal J}_b  b_{ij}^{\mu} b_{ij}^{\nu} + {\cal J}_c  c_{ij}^{\mu} c_{ij}^{\nu} 
\right) J_j^{\nu} \\
& &+  {\cal J}_4 \sqrt{2}~\vec{b}_{ij}.(\vec{J}_i \times \vec{J}_j)
\end{eqnarray*}
Considering for instance the pair of \er\, ions at $\vec r_1=(1/4,3/4,0)a$ and $\vec r_2=(0,1/2,0)a$, 
where $a$ is the cubic lattice constant, we define the local bond frame as:  $\vec{a}_{12} = (0,0,-1)$, 
$\vec{b}_{12} = 1/\sqrt{2} (1,-1,0)$ and $\vec{c}_{12} = 1/\sqrt{2} (-1,-1,0)$. This Hamiltonian, 
written in terms of bond-exchange constants, has the great advantage to provide a direct physical 
interpretation of the different parameters.

The magnetization is given by $M(\vec{H}) = \sum_i \vec{m}_i.\frac{\vec{H}}{H}$, where the 
$\vec{m}_i$ denote the individual magnetic moments. To carry out this calculation we assume a 
$\bf{k}$=0 magnetic structure in the Fd-3m space group with face centered cubic (fcc) symmetry. 
In other words, the 4 \er\, moments of a given tetrahedron may be different, but the spin 
configurations on tetrahedra connected by fcc lattice translations are the same. Following a 
self-consistent treatment, $\cal H$ is diagonalized numerically for each site to determine the 
energies $E_{i,\mu}$ and the wave functions $\vert \psi_{i,\mu} \rangle$. This yields the 
magnetic moment (see also Appendix \ref{MF_model}): 
\begin{equation*}
 \vec{m}_i=-g_J \mu_B \langle \vec{J}_i \rangle = 
-g_J \mu_B \sum_{\mu} \frac{e^{-E_{i,\mu}/k_B T}}{Z} \langle  \psi_{i,\mu}  | \vec{J}_i | \psi_{i,\mu}  \rangle
\end{equation*}
where $Z=\sum_{\mu}\exp{-E_{i,\mu}/k_B T}$. For a given field amplitude, this procedure 
is repeated for different directions to account for the powder average.

Since ${\cal J}_4$ is an anti-symmetric exchange constant (Dzyaloshinskii-Moriya like), it is expected to 
be smaller than the symmetric ones ${\cal J}_{a,b,c}$. Assuming ${\cal J}_4 = \pm 0.005$ K \cite{footnote2}, 
the magnetization curve is then well reproduced by the blue and black sets of parameters in Figure \ref{fig2}: 
\begin{eqnarray*}
{\cal J}_a \sim 0.03 \pm 0.017~{\rm K} \quad {\cal J}_b \sim 0.03 \pm 0.005 ~{\rm K}\\
{\cal J}_c \sim 0.04 \pm 0.005~{\rm K}
\end{eqnarray*}
Incorporating the nearest neighbors contribution of the dipolar interaction ($D_{nn}$=0.022 K, see 
Appendix \ref{MF_model}) in these anisotropic exchange constants leads to the effective parameters 
${\cal J'}_a \sim 0.05 \pm 0.017~{\rm K}$, ${\cal J'}_b \sim 0.05 \pm 0.005 ~{\rm K}$ and 
${\cal J'}_c \sim \pm 0.005~{\rm K}$. 

Next, it is of great interest to compare these results with the \erti\, exchange parameters listed in table 
\ref{table} and obtained from spin-waves \cite{savary} or magnetization curve analysis \cite{bonville}. 
We first note that the ${\cal J}_4$ value in \erti\, is also almost zero when considering the error bars. 
Interestingly, the symmetric exchange couplings in \ersn\, are smaller and more isotropic than in \erti, 
thus making the dipolar interaction the main anisotropic interaction. 

\begin{table}
\begin{tabular}{*{4}{c}}
\hline
Coupling & \ersn & \erti & \erti \\
& present work & Ref. \onlinecite{savary}  & Ref. \onlinecite{bonville}  \\ 
\hline
${\cal J}_a$  & 0.03 ($\pm$ 0.017) & -0.078 ($\pm$ 0.06)   & -0.030 ($\pm$ 0.01)   \\
${\cal J}_b$  & 0.03 ($\pm$ 0.005) & 0.078 ($\pm$ 0.01)    & 0.05 ($\pm$ 0.005)  \\
${\cal J}_c$  & 0.04 ($\pm$ 0.005) & 0.078 ($\pm$ 0.07)    & 0.105 ($\pm$ 0.01)    \\
${\cal J}_4$  & $\pm$ 0.005          & 0.02  ($\pm$ 0.03)     & $\pm$ 0.005  \\
$g_{\perp}$ & 7.52   & 5.97 & 6.8  \\
$g_ {//}$     & 0.054 & 2.45 & 2.6  \\
\hline
\end{tabular}
\caption{Anisotropic exchange parameters for \ersn\, (present work) and \erti\, \cite{savary, bonville}. 
Error bars are given in parenthesis. Positive values correspond to antiferromagnetic couplings. The 
conversion from original values \cite{savary} to the ${\cal J}_{a,b,c,4}$ set is detailed in Appendix 
\ref{pseudo-spin}.}
\label{table}
\end{table}

\begin{figure}[t]
\centerline{
\includegraphics[width=8cm]{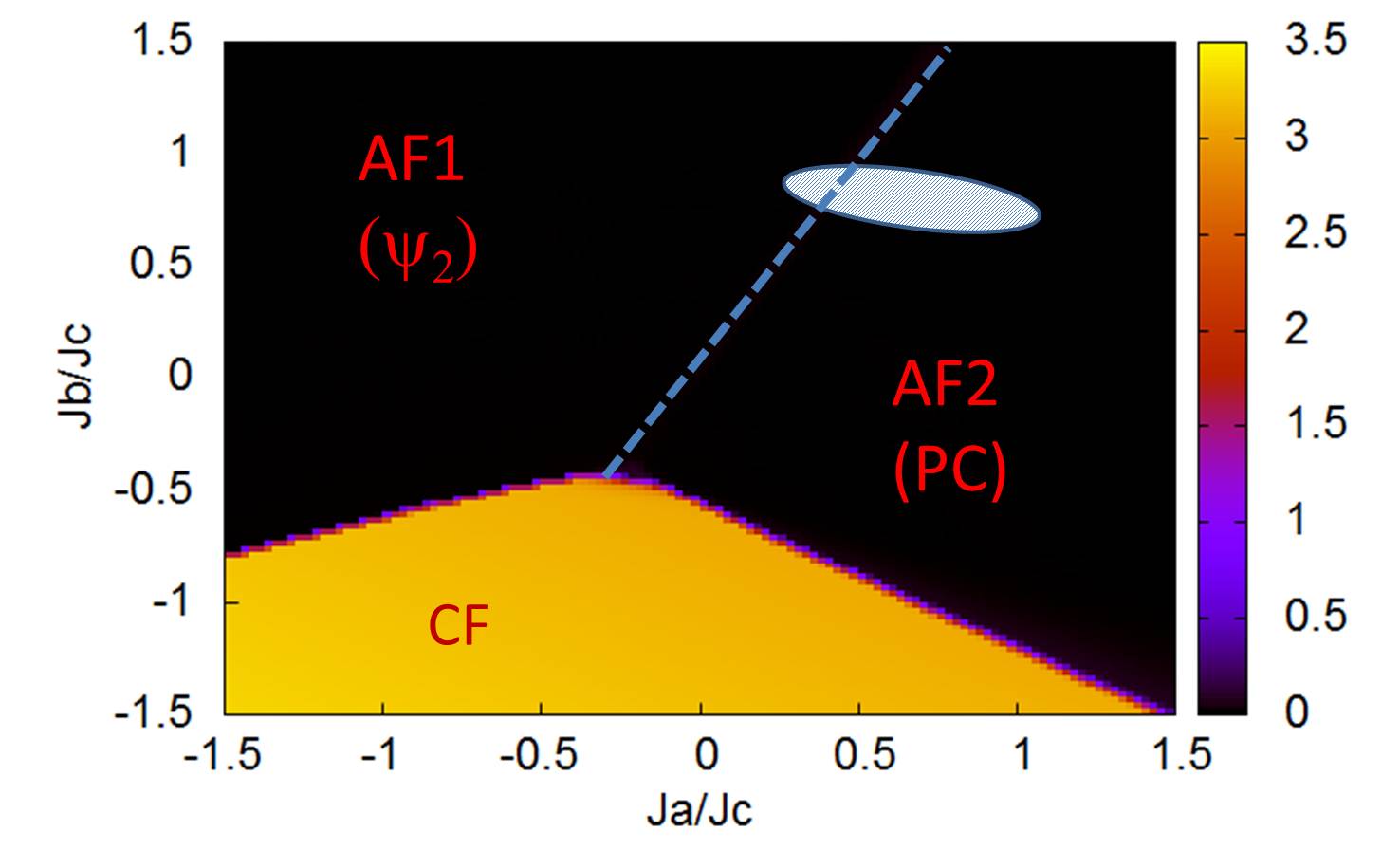}
}
\caption{(Color online) Mean field phase diagram for \ersn, (with the above determined CEF coefficients).
${\cal J}_c$ and ${\cal J}_4$ are fixed to 0.04 K and 0 respectively, while dipolar interaction is included. 
The AF1 phase resembles very much the $\psi_2$ state with the moments coordinates at the 4 sites 
(x,x,y), (-x,-x,y), (-x,x,-y), (x,-x,-y) and y$\approx$2x. The moments coordinates in the  CF phase are 
(x,x,y), (-x,-x,y), (x,-x,y), (-x,x,y). The shaded area corresponds to the region that accounts for the 
$M(H)$ measurements at 500 mK in \ersn. 
}
\label{fig5}
\end{figure}

In this context, according to a number of theoretical works \cite{clarty, stasiak, palmer}, the ground 
states in \ersn\, should belong to the $\Gamma_7$ representation, that is to say to the Palmer-Chalker 
states. The mean-field phase diagram (see Figure \ref{fig5}) computed in zero field as a function of 
${\cal J}_{a}/{\cal J}_{c}$ and ${\cal J}_{b}/{\cal J}_{c}$ confirms this assumption. As quoted in Ref.
\onlinecite{clarty}, the energetic selection at play in this approach is quite weak and neglects the influence 
of quantum and thermal fluctuations. Nonetheless, it is useful to explore the type of correlations that 
might develop depending on the exchange parameters. First, the calculation predicts a canted ferromagnetic
 "CF" state in the negative ${\cal J}_b/{\cal J}_c$ region, which might be relevant in the case of other 
XY pyrochlores, namely \ybti \cite{gingrasrmp, Chang12} or \ybsn \cite{Yaouanc13}. As for the \erti\, 
parameters, they lead in this phase diagram to a long-range ordered antiferromagnetic phase labeled 
AF1, almost identical to $\psi_2$ ($\Gamma_5$). This ground state is obtained for a strongly 
anisotropic exchange tensor and especially for ferromagnetic and weakly antiferromagnetic 
${\cal J}_a/{\cal J}_c$. Finally, the \ersn\, parameters lead to a different ground state labeled AF2 
which exactly corresponds to the Palmer-Chalker states ($\Gamma_7$), with a magnetic moment of 
3.9 $\mu_B$ and a mean-field ordering temperature $T_N \sim$ 1.3 K.

The obtained energy difference between the three states of $\Gamma_7$ is very small so that the ultimate 
selection is expected to be very fragile with respect to any fluctuations. This mean-field phase diagram thus 
confirms that the anisotropy and exchange parameters in \ersn\, stabilize Palmer-Chalker correlations as 
measured experimentally, suggesting that \ersn\, is akin to an XY pyrochlore antiferromagnet with dipolar 
interactions. 

The role of the latter interactions in stabilizing Palmer-Chalker states was  pointed out in the case of the 
Heisenberg pyrochlore \gdsn. It undergoes a first order transition towards a long range order at about 
1 K \cite{bonville03, wills06, cepas,stewart}, this ordering being robust with respect to quantum 
fluctuations \cite{delmaestro}. The lack of ordering in \ersn\, thus remains puzzling, but the XY 
anisotropy as well as the proximity of the AF1 phase might be key ingredients to explain it.
 
\section{Summary}

In summary, \ersn\, does not exhibit long range order down to the base temperature probed of 100 mK, but 
shows a macroscopic freezing below 200 mK. The magnetic moments have a very strong planar character. 
From the analysis of the magnetization curve within a mean-field model, the exchange couplings are found 
to be relatively weak and isotropic. At 1.5 K, the diffuse magnetic scattering is well reproduced by considering 
short-range correlations corresponding to Palmer-Chalker configurations. These results can be accounted for 
by a mean-field model which confirms that, with the \ersn\, parameters deduced from the experiments, 
Palmer-Chalker configurations, stabilized by the dipolar interactions, should be the ground state. In that 
context, the absence of ordering in \ersn\, remains an open issue, but the present estimation of the CEF 
parameters and exchange couplings appears to be a starting point for further theoretical calculation. 

\acknowledgements{We would like to acknowledge F. Damay and G. Andr\'e for comments about the use 
of the Fullproff suite as well as M. Gingras, B. Canals and M. Zhitomirsky for fruitful discussions. We thank 
C. Paulsen for allowing us to use his SQUID dilution magnetometers.}



\appendix

\section{Crystal field}
\label{CEF}
The CEF parameters are determined using the standard Hamiltonian. 
\begin{equation*} 
{\cal H}_{\mbox{CEF}} =
\sum_{m,n} B_{nm} O_{nm}
\end{equation*}
The transition between levels give rise to dispersionless modes in elastic neutron 
scattering data. The positions and intensities of these modes are fitted in the present 
study. To illustrate the sensitivity of our determination, we present in table \ref{table0} 
the results of CEF calculations for different sets of $B_{nm}$. The g-Land\'e factors are 
determined by considering the projection of the magnetic moment operator in the 
subspace spanned by the ground doublet wavefunctions. The error bars on the 
g-Land\'e factors are estimated from these calculations. We provide the energies of the two 
first transitions (experimentally observed at $E_1$=5.1 $\pm$ 0.05 and $E_2$=7.6 
$\pm$ 0.05 meV) as well as $g_{\perp}$ and $g_{//}$.\\

\begin{table}
\begin{tabular}{cccccccccc}
\hline
$B_{20}$ & $B_{40}$ & $B_{43}$ & $B_{60}$ & $B_{63}$ & $B_{66}$ & $E_1$ & $E_2$ &  $g_{//}$ & $g_{\perp}$\\
(K) & (K)  & (K)  & (K) & (K) & (K) & meV & meV & & \\
\hline
656	& 3010	& 755 &738	& -653	& 990  &5.08      &7.64    &0.054  &7.53\\
700	&		&	   &		&           &        &5.03	&7.70    &0.125 &7.57\\
600	&		& 	   &		&           &        &5.15	&7.56	&0.034	&7.47\\
	&2710	&	   &		&           &        &5.47	&8.59	&1.065	&7.81\\
	&3310	&	   &		&           &        &4.77	&6.92	&1.405	&7.09\\
	&	      &800  &		&           &        &4.90	&7.41	&0.013	&7.52\\
	&	      &700  &		&           &        &5.30	&7.91	&0.10	&7.53\\
	&  	      &	   & 800	&           &        &5.74	&7.91	&0.052	&7.47\\
	& 	      &	   & 700	&           &        &4.70	&7.48	&0.16	&7.57\\
	&	      &	   &	       &-600 	&        &4.93	      &7.18	&0.98	&7.28\\
	&	      &  	   &	       &-700	 &	    &5.24	      &8.14	&0.84	&7.68\\
	&	      &	   &	       &	       &1100 &5.38 	&8.64	&0.22	&7.64\\
	&	      &	   &	       &	       &900   &4.81 	&6.75	&0.17	&7.39\\
\hline
\end{tabular}
\caption{CEF calculations for different $B_{nm}$ parameters : the two first transitions 
$E_{1,2}$ along with the g-Land\'e factors $g_{\perp}$ and $g_{//}$ are listed.}
\label{table0} 
\end{table}

\section{Details about the $\psi_2$ basis vector}
\label{mag_struc}

The description of the possible magnetic structures in \erti\, and \ersn\, XY 
antiferromagnets is based on the symmetry analysis performed in the Fd-3m space 
group for a ${\bf k}=0$ 
propagation vector. As explained in the main text, \erti\, undergoes a transition towards 
an antiferromagnetic N\'eel phase below $T_N=1.2$~K \cite{Blote69, Harris98, Siddharthan99}. 
This ordered phase has a non-collinear structure, in which the magnetic moments are 
perpendicular to the local $<111>$ axes. This configuration corresponds to the $\psi_2$ 
basis vector of the $\Gamma_5$ irreducible representation.

To better figure out this peculiar configuration, a different approach can be followed 
\cite{savary,zito}, considering a series of XY degenerate classical configurations where the 
magnetic moment at site i is defined in a local frame $(\vec{a}_i,\vec{b}_i,\vec{e}_i)$ 
given in Table \ref{table1}. Each magnetic moment points along 
$\vec{u}_i = \cos{\phi}~\vec{a}_i + \sin{\phi}~\vec{b}_i$, where $\phi$ is a continuous 
parameter. $\vec{e}_i$ is the local CEF axis.
With these notations, the 6 domains of the $\psi_2$ magnetic structure are obtained for 
$\phi=n \pi/3, n=0,..,5$, while the $\psi_1$ (Ref \onlinecite{savary,champ2}, also called 
$\psi_3$ in Ref. \onlinecite{zito,poole}) are generated for $\phi=\pi/6+n \pi/3, n=0,..,5$.

These configurations are classically degenerate since for arbitrary $\phi$, the classical energy 
given by $E_c= (4 {\cal J}_a-2 (3 {\cal J}_b+{\cal J}_c)) m^2$, where 
$({\cal J}_a,{\cal J}_b,{\cal J}_c)$ are the bond exchange parameters defined in the main text, 
does not depend on $\phi$. The studies published in \cite{savary,zito} have shown further that 
the zero point energy $E_o(\phi)$, calculated in the spin wave approximation as a function of $\phi$, 
breaks this degeneracy, exhibiting weak minima for the 6 $\psi_2$ domains. A particular ordered 
ground state is thus selected by this quantum order by disorder mechanism.\\

\begin{table}[h]
\begin{tabular}{ccccc}
\hline
Site & 1 & 2 & 3 & 4 \\
CEF axis $\vec{e}_i$ & (1,1,-1) & (-1,-1,-1) & (-1,1,1) & (1,-1,1) \\
\hline
Position          & ($\frac{1}{4}$, $\frac{3}{4}$, 0) & (0, $\frac{1}{2}$, 0) & (0, $\frac{3}{4}$,$\frac{1}{4}$) & ($\frac{1}{4}$, $\frac{1}{2}$, $\frac{1}{4}$) \\
$\vec{a}_i$       & (-2,1,-1) & (2,-1,-1) & (2,1,1) & (-2,-1,1) \\
$\vec{b}_i$       & (0,1,1) & (0,-1,1) & (0,1,-1) & (0,-1,-1) \\
\hline
\end{tabular}
\caption{$(\vec{a}_i,\vec{b}_i,\vec{e}_i)$ frame for the different sites of a tetrahedron.}
\label{table1}
\end{table}

\section{Mean field model}
\label{MF_model}
The present mean field study follows the approach of Ref \onlinecite{clarty}; it is based 
on the following Hamiltonian for R moments $\vec{J}_i$ at site $i$:
\begin{equation*}
{\cal H} = {\cal H}_{\mbox{CEF}} + 
\sum_{<i,j>} \vec{J}_i ({\cal \tilde J}+{\cal \tilde J}\mbox{dip}_{i,j})
\langle \vec{J}_j \rangle + g_J \mu_B \vec{H}. \vec{J}_i
\end{equation*}
In this expression, $\vec{H}$ is an applied magnetic field, ${\cal \tilde J}$ denotes the anisotropic 
exchange tensor and ${\cal \tilde J}\mbox{dip}_{i,j}$ the dipolar interaction limited to the contribution 
of the nearest neighbours. Various conventions have been used to define this anisotropic exchange 
\cite{clarty,thompson,savary,zito}. 
Here, we assume an exchange tensor which is diagonal in the $(\vec{a},\vec{b},\vec{c})$ frame linked 
with a R-R bond. Considering for instance the pair of \er\, ions at $r_1=(1/4,3/4,0)a$ and $r_2=(0,1/2,0)a$, 
where a is the cubic lattice constant, we define the local bond frame as: $\vec{a}_{12} = (0,0,-1)$, 
$\vec{b}_{12} = 1/\sqrt{2} (1,-1,0)$ and $\vec{c}_{12} = 1/\sqrt{2} (-1,-1,0)$. 
\begin{eqnarray*}
\vec{J}_i {\cal \tilde J} \vec{J}_j &=& \sum_{\mu,\nu=x,y,z} J_i^{\mu} 
\left( 
{\cal J}_a a_{ij}^{\mu} a_{ij}^{\nu} +  
{\cal J}_b  b_{ij}^{\mu} b_{ij}^{\nu} + {\cal J}_c  c_{ij}^{\mu} c_{ij}^{\nu} 
\right) J_j^{\nu} \\
& &+  {\cal J}_4 \sqrt{2}~\vec{b}_{ij}.(\vec{J}_i \times \vec{J}_j)
\end{eqnarray*}
Owing to the form of the dipolar interaction, we have :
\begin{equation*}
{\cal \tilde J}\mbox{dip}_{i,j} = D_{nn} \left(\vec{a}_{ij}\vec{a}_{ij} + \vec{b}_{ij}\vec{b}_{ij} - 2 \vec{c}_{ij}\vec{c}_{ij} \right)
\end{equation*}
with $D_{nn} = \frac{\mu_o}{4\pi}\frac{(g_J \mu_B)^2}{r^3_{nn}}$ and where $r_{nn}$ is the 
nearest neighbour distance in the pyrochlore lattice. If we incoporate it in the anisotropic exchange 
constants $({\cal J}_a,{\cal J}_b,{\cal J}_c)$, we obtain : 
\begin{eqnarray*}
{\cal J'}_a  &=& {\cal J}_a + D_{nn} \\
{\cal J'}_b  &=& {\cal J}_b + D_{nn} \\ 
{\cal J'}_c  &=& {\cal J}_c - 2  D_{nn}
\end{eqnarray*}
As usual in mean field approximations, a self-consistent treatment is carried out to solve the 
problem: starting from a random configuration for the $\langle \vec{J}_j \rangle$, the 
contribution to $\cal H$ at site $i$ is diagonalized in the Hilbert space of the \er\, magnetic 
moment defined by the $\left\{ | J_z \rangle \right\}, J_z=-15/2,...,15/2$ basis vectors, and 
taking into account the external magnetif field $\vec{H}$ as well as the molecular field $
\sum_{<i,j>} \vec{J}_i ({\cal \tilde J}+{\cal \tilde J}\mbox{dip}_{i,j})
\langle \vec{J}_j \rangle$. This yields the energies $E_{i,\mu}$ and the wave functions 
$\vert \psi_{i,\mu} \rangle$. The updated magnetic moments: 
\begin{eqnarray*}
 \langle \vec{J}_i  \rangle & =
&  \sum_{\mu} \frac{e^{-E_{i,\mu}/k_B T}}{Z} \langle  \psi_{i,\mu}  | \vec{J}_i | \psi_{i,\mu} \rangle \\
Z & = & \sum_{\mu} \exp{-E_{i,\mu}/k_B T}
\end{eqnarray*}
is used to proceed at site $j$, and this is repeated until convergence. 
The magnetization is then given by: 
\begin{equation*}
M(\vec{H}) = \sum_i  \vec{m}_i.\frac{\vec{H}}{H}
\end{equation*}

\begin{table}[t]
\begin{tabular}{*{4}{c}}
\hline
Coupling & \erti & \erti & \ersn \\
& Ref \onlinecite{savary}  & Ref \onlinecite{bonville} & present work \\ 
\hline
${\sf J}_{\pm \pm}$    & 4.2 ($\pm$ 0.5)     & 3.2 ($\pm$ 1)     & 7.4 ($\pm$ 1.5)\\
${\sf J}_{\pm}$           & 6.5 ($\pm$ 0.75)   & 6.7 ($\pm$ 1)     & 1.35 ($\pm$ 1.5)\\
${\sf J}_{z\pm}$          & -0.88 ($\pm$ 1.5)  & 1.32 ($\pm$ 0.5)    & 0.025 ($\pm$ 0.01)\\
${\sf J}_{zz}$              & -2.5 ($\pm$ 1.8)    & -1.75 ($\pm$ 0.4)  & 0.0 \\
\hline
${\cal J}'_a$ & -0.056 ($\pm$ 0.06)   & -0.008 ($\pm$ 0.01)     & 0.052 ($\pm$ 0.017) \\
${\cal J}'_b$ & 0.10  ($\pm$ 0.01)     & 0.072   ($\pm$ 0.005)  & 0.052 ($\pm$ 0.005) \\
${\cal J}'_c$ & 0.034 ($\pm$ 0.07)     & 0.061   ($\pm$ 0.01)   & -0.004 ($\pm$ 0.005) \\
${\cal J}_4$ & 0.02  ($\pm$ 0.03)      & $\pm$ 0.005               & 0 \\   
\hline
${\cal J}_a$ & -0.078 ($\pm$ 0.06)     & -0.03 ($\pm$ 0.01) & 0.03 ($\pm$ 0.017) \\
${\cal J}_b$ & 0.078  ($\pm$ 0.01)     & 0.05 ($\pm$ 0.005)  & 0.03 ($\pm$ 0.005) \\
${\cal J}_c$ & 0.078 ($\pm$ 0.07)      & 0.105 ($\pm$ 0.01)    & 0.04 ($\pm$ 0.005) \\
${\cal J}_4$ & 0.02  ($\pm$ 0.03)      & $\pm$ 0.005            & $\pm$ 0.005 \\
\hline
$g_{\perp}$ & 5.97 & 6.8 & 7.52 \\
$g_ {//}$     & 2.45 & 2.6 & 0.054 \\
\hline
\end{tabular}
\caption{Anisotropic exchange parameters for \erti\, and \ersn. 
(${\sf J}_{\pm \pm}$, ${\sf J}_{\pm}$, ${\sf J}_{z\pm}$,${\sf J}_{zz}$) are given in 
$10^{-2}$ meV while the other sets are in K. Positive values correspond to AF couplings. }
\label{table-param}
\end{table}
\section{Relation with quantum pseudo-spin half models}
\label{pseudo-spin}
The anisotropic exchange Hamiltonian can be rewritten in terms of couplings between the spin 
components of a pseudo spin half defined in the subspace of the ground CEF doublet:
\begin{eqnarray*}
{\cal H}' &=& \sum_{i,j} {\sf J}_{zz} {\sf S}^z_i {\sf S}^z_j - {\sf J}_{\pm} \left({\sf S}^+_i {\sf S}^-_j + {\sf S}^-_i {\sf S}^+_j \right) \\
& &
+ {\sf J}_{\pm\pm} \left(\gamma_{ij} {\sf S}^+_i {\sf S}^+_j + \gamma^*_{ij} {\sf S}^-_i {\sf S}^-_j \right) \\
& &
+ {\sf J}_{z \pm} \left[ {\sf S}_i^z \left( \zeta_{ij} {\sf S}^+_j + \zeta^*_{ij} {\sf S}^-_j\right) + i \leftrightarrow j \right] 
\end{eqnarray*} 
$({\sf J}_{\pm\pm},{\sf J}_{\pm},{\sf J}_{z\pm},{\sf J}_{zz})$ is the set of effective exchange parameters. 
Note that "sanserif" notations refer to local bases. The states of this pseudo spin half span the ground CEF 
wavefunctions doublet, using the relation :
\begin{equation}
g_J \vec{{\sf J}} = g \vec{{\sf S}}~~~\mbox{or}~~~~ \vec{{\sf J}} = \lambda \vec{{\sf S}}
\end{equation}
In the context of pyrochlores, the $\lambda=\frac{g}{g_J}$ matrix is diagonal and takes the form:
\begin{equation}
\lambda = \left(
\begin{array}{ccc}
\lambda_{\perp} & & \\
 & \lambda_{\perp} & \\
& & \lambda_z
\end{array}
\right)
\end{equation}

We call $M$ the matrix that connects the local and global bases and $A$ the matrix 
connecting $({\sf S_x},{\sf S_y},{\sf S_z})$ and $({\sf S_+},{\sf S_-},{\sf S_z})$ (we omit 
the indexes for sake of clarity), so that:
\begin{equation}
\vec{J} = M~ \lambda ~A~\vec{{\sf S}}
\end{equation}
\begin{equation}
A = \left(
\begin{array}{ccc}
1/2 & 1/2& \\
-i/2& i/2 & \\
& & 1
\end{array}
\right)
\end{equation}
with $\vec{{\sf S}}=({\sf S_+},{\sf S_-},{\sf S_z})$. 
We thus have :
\begin{equation}
{\cal H} = \sum_{ij,uv} {\sf S}_i^u~\left( A^T ~\lambda~M^T_i ~{\cal J}~ M_j~ \lambda ~A \right)^{uv}~{\sf S}_i^v
\end{equation}
and we finally get the following relations:
\begin{eqnarray*}
{\sf J}_{zz}      & = & \lambda_z^2 ~\frac{{\cal J}_a-2{\cal J}_c-4{\cal J}_4}{3} \\
{\sf J}_{\pm}    & = & -\lambda_{\perp}^2 ~\frac{2{\cal J}_a-3{\cal J}_b-{\cal J}_c+4{\cal J}_4}{12} \\
{\sf J}_{z\pm}   & = & \lambda_{\perp}~\lambda_z ~\frac{{\cal J}_a+{\cal J}_c-{\cal J}_4}{3 \sqrt{2}} \\
{\sf J}_{\pm \pm} & = & \lambda_{\perp}^2 ~\frac{2{\cal J}_a+3{\cal J}_b-{\cal J}_c+4{\cal J}_4}{12}
\end{eqnarray*}
and conversely:
\begin{eqnarray*}
{\cal J}_a & = & \frac{4}{3}~\frac{{\sf J}_{\pm\pm} - {\sf J}_{\pm }}{\lambda_{\perp}^2} + \frac{4 \sqrt{2}}{3} \frac{{\sf J}_{z\pm}}{\lambda_{\perp}\lambda_z} + \frac{1}{3} \frac{{\sf J}_{zz}}{\lambda_z^2} \\
{\cal J}_b & = & 2~\frac{{\sf J}_{\pm\pm} + {\sf J}_{\pm }}{ \lambda_{\perp}^2} \\
{\cal J}_c & = & \frac{2}{3}~\frac{-{\sf J}_{\pm\pm} + {\sf J}_{\pm }}{ \lambda_{\perp}^2} + \frac{4 \sqrt{2}}{3} \frac{{\sf J}_{z\pm}}{\lambda_{\perp}\lambda_z} - \frac{2}{3} \frac{{\sf J}_{zz}}{\lambda_z^2} \\
{\cal J}_4 & = & \frac{2}{3}~\frac{{\sf J}_{\pm\pm} - {\sf J}_{\pm }}{ \lambda_{\perp}^2} - \frac{\sqrt{2}}{3} \frac{{\sf J}_{z\pm}}{\lambda_{\perp}\lambda_z} - \frac{1}{3} \frac{{\sf J}_{zz}}{\lambda_z^2}
\end{eqnarray*}

Table \ref{table-param} provides the different sets of anisotropic exchange parameters for \erti\, (Ref. \onlinecite{savary,bonville}) and \ersn\, (present work) deduced from this transformation. This 
procedure is similar to the ones detailed in references \cite{thompson,savary,curnoe}. 

\begin{thebibliography}{99}
\bibitem{neutron} On 4F2, we used a fixed final energy ($E_f$ = 5 meV), yielding an instrumental resolution of 
about 0.25 meV. A nitrogen-cooled beryllium filter placed in the scattered beam to eliminate harmonics.
\bibitem{notation} Here, we adopt the notations of \cite{champ2, savary}; note that they are different
from those used in \onlinecite{poole} and \onlinecite{zito}.
\bibitem{footnote1} The sample being a powder, the demagnetizing factor $N$ is hard to estimate. 
However, the main characteristics of the exchange tensor deduced from the refinement of the magnetization 
curve are only slightly affected by the chosen value of $N$, so that the present analysis does not depend on its 
exact value.
\bibitem{footnote2} Fits were performed setting ${\cal J}_4$ to a series of values between -0.02 and 
+0.02 (which is about 50\% of the symmetric constants).  For ${\cal J}_4$ larger than 0.01, the 
magnetization curve could not be reproduced. Note that, whatever ${\cal J}_4$, all the sets of 
parameters that reproduce the magnetization curve fall in the Palmer-Chalker part of the mean-field 
phase diagram of Figure \ref{fig5}.
\end{thebibliography}
\end{document}